\documentclass[11pt]{article}
\usepackage{moriond,epsfig}
\usepackage{amssymb,amsmath}

\def\be{\begin{equation}}
\def\ee{\end{equation}}
\def\bea{\begin{eqnarray}}
\def\eea{\end{eqnarray}}

\begin{document}
\vspace*{4cm}
\title{Overview of the heavy ion session: Moriond 2006}

\author{ Carlos A. Salgado}

\address{Dipartimento di Fisica, Universit\`a di Roma "La Sapienza"}

\maketitle\abstracts{We review recent theoretical ideas and experimental data in high-energy heavy ion collisions with special emphasis in the hard probes. 
}

Traditionally, high-energy physics experiments with colliders attempt to achieve the largest possible energy scales, in well localized spacial regions, for new physics to become observable. A new complementary direction is rapidly emerging in the last years which aims to produce the largest possible {\it extended} regions of high energy densities. The goal is, in this case, to understand how the collective properties of the fundamental interactions appear. At present and near-future reachable energies, the only theory which can be studied in such a conditions is QCD, and the relevant experiments are high-energy heavy ion collisions. The large spacial dimension of the ions provide the desired extension of the medium and large energy densities are produced at the center of mass of the collision. 

The relevant questions which can be addressed in experiments of heavy ion collisions can be (artificially) classified depending on the time scale of the relevant phenomena: i) before the collision, the structure of the two Lorenz-contracted nuclei is rather different from a typical hadron at the same energy and collective phenomena, nowadays generically known under the name of Color Glass Condensate, appear; ii) once the high-density state is created, the relevant question is how thermalization and other collective mechanisms, as a possible hydrodynamical behavior, appear; iii) finally, the properties of the eventually equilibrated medium need to be studied by means of some indirect signals. One main question in this last item is the relation with first principle calculations in QCD, namely from lattice, where different states of matter and possible phase transitions are predicted \cite{schmidt,levkova}.

Reaching collider energies provides the additional tool of the hard part of the spectrum, characterized by large virtualities, to be used for these studies. A typical hard cross section can be written in the form
\begin{equation}
\sigma^{AB\to h}=
f_A(x_1,Q^2)\otimes f_B(x_2,Q^2)\otimes \sigma(x_1,x_2,Q^2)\otimes D_{i\to h}
(z,Q^2)\, ,
\label{eqhard}
\end{equation}
with a factorization between the short-distance perturbative cross section $\sigma(x_1,x_2,Q^2)$, computable in powers of $\alpha_s(Q^2)$ and two long-distance terms, the proton/nuclear parton distribution functions (PDF), $f_A(x,Q^2)$, encoding the partonic structure of the colliding objects, and the fragmentation functions (FF), $D(z,Q^2)$, describing the hadronization of the parton $i$ into a final hadron $h$. These long-distance terms are non-perturbative objects, but whose evolution can be computed in perturbative QCD. These are precisely the objects which will be modified in the case that the {\it extension} of the colliding system interferes with the dynamics, while the short-distance part is expected to remain unchanged if the virtuality is large enough. It is this interference between geometry and dynamics which makes hard probes perfect tools to characterize the medium properties through the modification of the long-distance terms in (\ref{eqhard}).

A conceptually simple example is the case of the $J/\Psi$ whose production cross section is
\begin{equation}
\sigma^{hh\to J/\Psi}=
 f_i(x_1,Q^2)\otimes f_j(x_2,Q^2)\otimes
\sigma^{ij\to [c\bar c]}(x_1,x_2,Q^2)
 \langle {\cal O}([c\bar c]\to J/\Psi)\rangle\, ,
\end{equation}
where now $ \langle {\cal O}([c\bar c]\to J/\Psi)\rangle$ describes the hadronization of a $c\bar c$ pair in a given state (for example a color octet) into a final $J/\Psi$. This long-distance part is expected to be modified in a medium at finite temperature in which the deconfinement avoids the bound states to exist \cite{Matsui:1986dk}. However, this modification, being non-perturbative, lacks of good theoretical control and the real situation about the $J/\Psi$-suppression is complicated \cite{tram}. 

From the computational point of view, a theoretically simpler case involves the modification of the {\it evolution} of both the parton distribution and the fragmentation functions in a dense or finite--temperature medium. This needs of large scales $Q^2$ in order for the strong coupling to be small and perturbative methods to apply. In general, the presence of a dense medium is translated into non-linear terms in the evolution equations. 

\begin{figure}
\begin{minipage}{0.35\textwidth}
\begin{center}
\includegraphics[width=\textwidth]{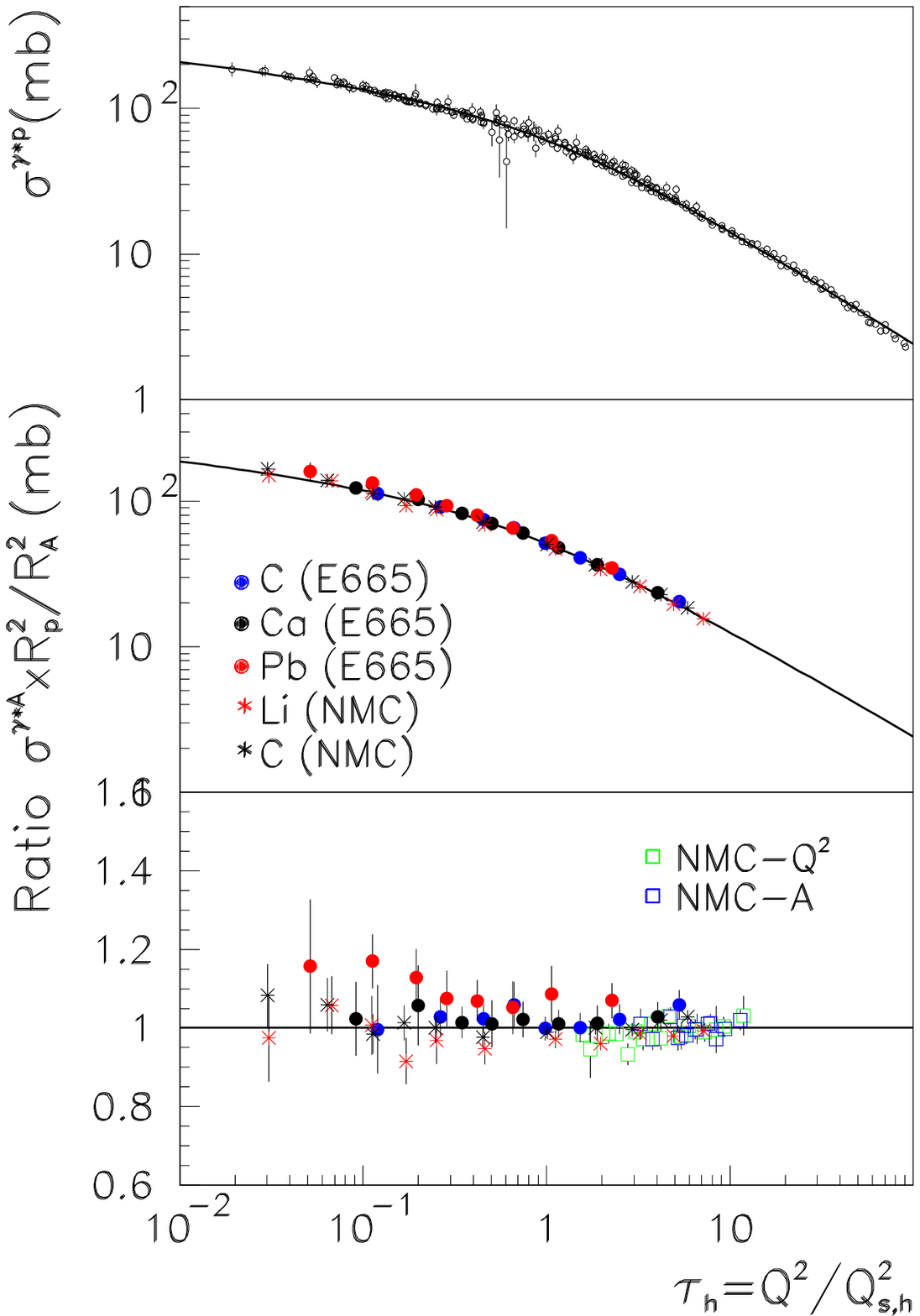}
\end{center}
\end{minipage}
\hfill
\begin{minipage}{0.65\textwidth}
\begin{center}
\includegraphics[width=0.8\textwidth]{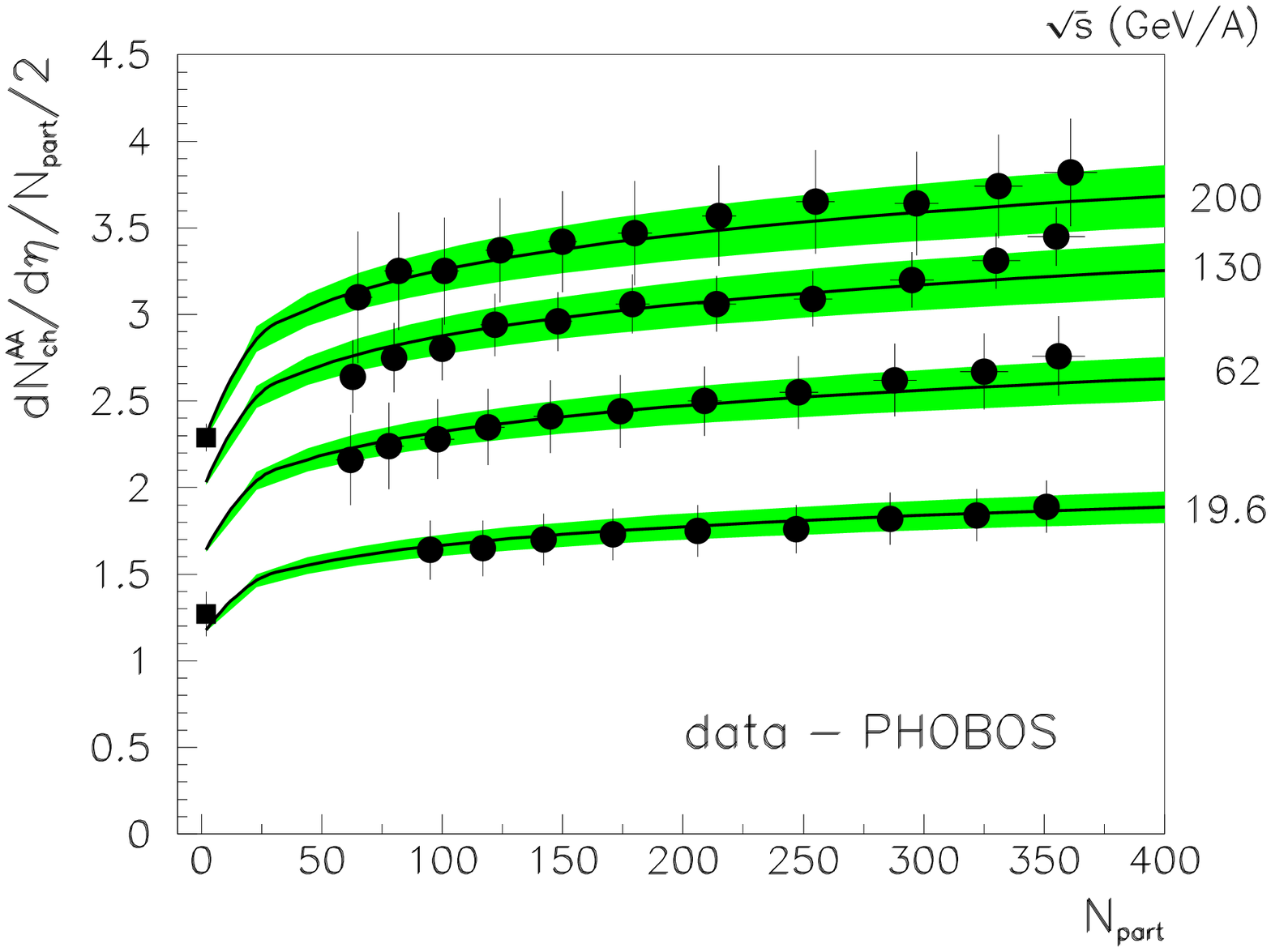}
\end{center}
\end{minipage}
\caption{Left: Geometric scaling in lepton-proton \protect\cite{Stasto:2000er} and lepton-nucleus \protect\cite{Armesto:2004ud} data. Right: Central rapidity multiplicities in $p\bar p$ and AuAu collisions at different centralities \protect\cite{loizides} and the corresponding description from Eq. (\ref{eqmult})\protect\cite{Armesto:2004ud}.}
\label{fig1}
\end{figure}

The study of such a modifications for the nuclear structure is known under the generic name of {\it Color Glass Condensate} which provides a general framework, based on an effective theory separating the fast modes from the {\it generated} slow modes, associated to small-$x$ gluons in the nuclear wave function. This small-$x$ gluons have parametrically large (${\cal O}(1/\alpha_s)$) densities and can be treated as classical fields. The quantum evolution equation of this setup is known and, remarkably, can be written in a rather simple form in the large-$N_C$ limit \cite{Iancu:2003xm}. A sophisticated technology has been developed in the last decade in this framework \cite{iancu,lublinsky,gelis} which, in addition to describe the structure of the incoming wave functions, aims to provide the link to the subsequent evolution into a thermal system \cite{Lappi:2006fp}. From a phenomenological point of view, the most successful applications of these formalism are the description of the multiplicities measured at RHIC \cite{Kharzeev:2000ph} -- see \cite{loizides} for an updated experimental status; the possible presence of the predicted geometric scaling in lepton-hadron data \cite{Stasto:2000er,Freund:2002ux,Armesto:2004ud}; and the suppression of inclusive particles at forward rapidities at RHIC \cite{Arsene:2004ux}, which has been predicted as a result of small-$x$ quantum evolution \cite{forwardsup}. A particularly economic description is presented in Fig. \ref{fig1}. Here, the saturation scale is obtained from lepton-proton and lepton-nucleus data as $Q^2_{\rm sat}\propto x^{-\lambda}A^{1/3\delta}$ with fitted parameters $\lambda=0.288$ and $\delta=0.79$. Fig. \ref{fig1} left shows the quality of the geometric scaling in lepton-proton \cite{Stasto:2000er} and lepton-nucleus \cite{Armesto:2004ud} data. Assuming the same scaling to hold in AA collisions, the multiplicity in the central rapidity can be written as \cite{Armesto:2004ud}
\begin{equation}
\frac{1}{N_{\rm part}}
\frac{dN^{AA}}{d\eta}\Bigg\vert_{\eta\sim 0}=N_0\sqrt{s}^\lambda
N_{\rm part}^{\frac{1-\delta}{3\delta}}\, .
\label{eqmult}
\end{equation}
where only a total normalization factor $N_0$ is needed once the energy and centrality dependences are fixed by lepton-proton and lepton-nucleus data respectively. Fig. \ref{fig1} shows the comparison of this simple formula with available data\cite{loizides}. Recent developments in this approach are the inclusion of some additional terms in the resummed series, known as pomeron-loops \cite{lublinsky,iancu}; the description in terms of statistical physics \cite{iancu}; the production of quarks and gluons in strong fields \cite{gelis}.

On the opposite side, a perturbative modification of the fragmentation of the highly virtual produced partons, has been proposed as a tool to characterize the medium produced in heavy ion collisions \cite{Salgado:2005pr}. These effects on high-$p_t$ particles, generically known as {\it jet quenching}, constitute one of the main experimental observations at RHIC so far \cite{Adcox:2001jp}. In the near future, the LHC will extend the range in transverse momentum for at least one order of magnitude. In the vacuum, a perturbatively produced high-$p_t$ parton with virtuality $Q^2\sim p_t^2$ develops a parton shower by radiating other partons (mainly gluons) with decreasing virtuality. This shower stops when a typical hadronic scale ${\cal O}(1 $GeV$^2)$ is reached. The resulting jet is an extended object in which the collinear and soft emitted gluons are emitted inside a cone around the original parton direction. When the high-$p_t$ particle is produced in the medium created after a heavy ion collision, this parton shower is modified. The jets, being extended structures, provide an excellent tool to characterize the medium properties at different scales. Indeed, the typical radiation time of a gluon with energy $\omega$ and transverse momentum $k_t$ is $t_{\rm form}\sim \omega/k_t^2$, which can be smaller than the medium size; the typical hadronization time for a bare parton to develop its hadronic wave function (boosted to the laboratory frame) is $t_{\rm had}\sim E/m\, R_{\rm had}$. Although, the concept of hadronization time is not clear in a deconfined medium, in most of the applications, one simply assumes that this time is large enough for the (non-perturbative) hadronization to take place in the vaccum, so that only the perturbative evolution needs to be modified.

At high enough parton energies, the main mechanism driving the modifications of high-$p_t$ evolution is the medium-induced gluon radiation. As in general high-energy processes, the propagation of the partons through the medium can be described in terms of Wilson lines averaged in the allowed configuration of a medium. Several prescriptions exists for these averages and in the multiple soft scattering limit, the saddle point approximation of the Wilson lines define a single parameter of the medium, the transport coefficient, $\hat q$, given by the average transverse momentum squared acquired by the gluon per mean free path. The typical energy spectrum of radiated gluons is softer than in the vacuum although formation time effects provide a tipical scale $\hat \omega\sim \hat q^{1/3}$ under which the radiation is suppressed. In the same way, the angular distribution is regulated at angles smaller than a typical angle $\hat\theta\simeq(\hat q/\omega^3)^{1/4}$. As a result of formation time effects (or coherence effects) the medium-induced gluon radiation is, hence, finite in the infrared or collinear limits. The two main predictions are the suppression of high-$p_t$ yields due to additional in-medium energy loss and the associated broadening of the jet structures.

\begin{figure}
\begin{center}
\includegraphics[width=0.43\textwidth,angle=-90]{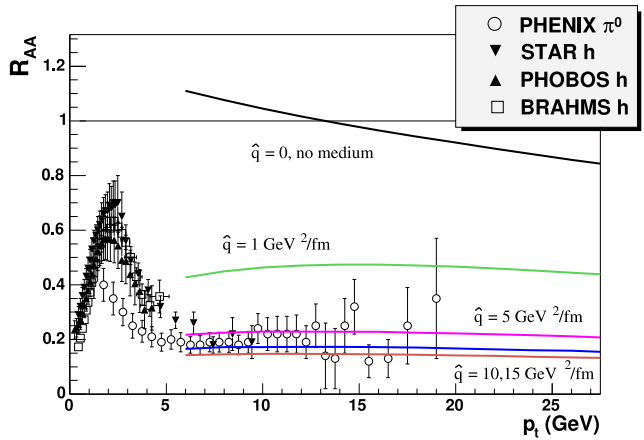}
\end{center}
\caption{Nuclear modification factor, $R_{AA}$, in central
AuAu collisions at $\sqrt{s}$=200 GeV \protect\cite{Eskola:2004cr}. Data from \protect\cite{Adcox:2001jp}.}
\label{figraa}
\end{figure}

Although the degree of theoretical refinement of the jet quenching formalisms is still not completely satisfactory, it provides a good description of experimental data with the medium-modified fragmentation function computed as
\begin{equation}
D_{i\to h}^{\rm med}(z,Q^2)=P_E(\epsilon)\otimes D_{i\to h}(z,Q^2)
\label{eqff}
\end{equation}
here, $P_E(\epsilon)$ gives the probability of an additional in-medium energy loss and is normally computed by assuming a simple Poisson distribution with the medium-induced gluon radiation as input. 
Once the geometry of the system is correctly taken into account, a fit to RHIC data (see Fig. \ref{figraa}) gives the value of the time-averaged transport coefficient $\hat q\sim 5...15$ GeV$^2$/fm. This large value and the corresponding uncertainty is a direct consequence of the surface trigger bias effect in inclusive particle suppression measurements \cite{Muller:2002fa,Eskola:2004cr,Dainese:2004te}. This is an intrinsic limitation of these type of measurements on the characterization of the medium and on the study of the dynamics underlying the propagation of highly energetic partons through a dense medium. Further constraints can be found by i) measuring different particles species, and in particular heavy quarks, as the formalism predicts the hierarchy $\Delta E_g>\Delta E_q^{\rm m=0}>\Delta E_Q^{\rm m\neq 0}$; ii) by directly measuring the induced radiation, i.e. by reconstructing the jet structure in a heavy ion collision.

\begin{figure}
\begin{center}
\includegraphics[width=0.36\textwidth,angle=-90]{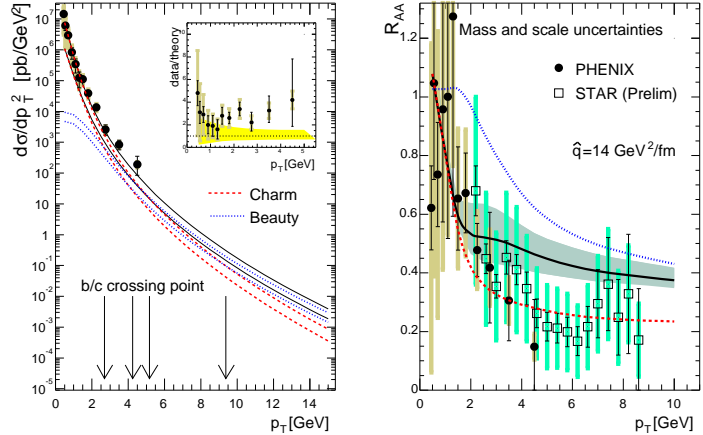}
\end{center}
\caption{Left: Comparison of the FONLL calculation of single 
inclusive electrons from pp collisions at $\sqrt{s}$=200GeV \protect\cite{Cacciari:2005rk}. Right: The nuclear modification factor of electrons with the corresponding uncertainty coming from the perturbative benchmark on the relative $b/c$ contribution. Figure from \protect\cite{Armesto:2005mz}; data from \protect\cite{elqm05}.}
\label{fighq}
\end{figure}

New data from RHIC on non-photonic electrons \cite{elqm05} attempt to answer the question on heavy quark in-medium energy loss. This electrons are expected to come from the decays of charm and beauty quarks. The perturbative description of the relative contribution of both quarks to the electron yield is, however, not under good control and the $b/c$ crossing point can be as low as 2 GeV or as large as 10 GeV when the usual mass and scale uncertainties are taken into account in   approximation. This translates into a large uncertainty in the ratio $R_{AA}^e$ as can be seen in Fig. \ref{fighq} \cite{Armesto:2005mz}. The description of the experimental data is reasonable within the error bars, although not completely satisfactory. A clear distinction between heavy mass effects in medium-modified gluon radiation seems only possible with a better identification of the $c$ and $b$ contributions and, ideally, by a direct measurement of the $D$ and $B$ mesons. This identification would constrain the importance of other mechanisms of heavy quark energy loss \cite{Casalderrey-Solana:2006rq}.

The most direct probe of the validity of the medium-induced gluon radiation as the underline dynamics for jet quenching would be the modification of jet shapes. Two main uncertainties appear in the callibration of the jet energy, on the one hand, out-of-cone fluctuations, reducing the measured energy of the jet, increase with decreasing cone opening angle $R=\sqrt{\Delta\eta^2+\Delta\phi^2}$; on the other hand, background fluctuations, increasing the measuring jet energy, increase with increasing $R$. Although these uncertainties are common to any colliding system, in a situation of large multiplicity background, as it is the case in heavy-ion collisions, they become the real issue to overcome \cite{belikov}. From a theoretical side, the identification of observables with small sensitivity to background substractions is needed \cite{Salgado:2003rv}. 

Although the measurement of true jets in heavy ion collisions will be only possible at the LHC, with much larger yields and accessible transverse momentum, data from two-particle correlations at RHIC is already reveling the richness of this physics. In particular, for the first time measurements of two-particle azimuthal correlations at large transverse momentum, with negligible combinatorial background, have been recently released \cite{Adams:2006yt}. These data shows some unexpected features, namely, the absence of broadening in the direction opposite to the trigger particle (away-side direction) and a strong suppression of this same yield, with a magnitude similar to that for the inclusive yields. A natural interpretation in terms of the jet-quenching formalism is that this two-particle inclusive measurement choses those event characterized by small in-medium pathlength, associated with small energy losses and, hence, small broadening \cite{Dainese:2004te}. When the transverse momentum of the associated particles is lowered, interesting effects appear, in particular a very large broadening of the away-side signal, with the possible appearance of two peaks with a dip at $\Delta\phi=\pi$ \cite{Adler:2005ee}. The evidence for these type of structures is not yet conclusive, it depends strongly on the $p_t$ cuts and the background subtraction and different experiments do not completely agree. However, it is stimulating new theoretical ideas about the interplay between the soft-bulk and the hard part of the spectrum \cite{Armesto:2004pt,conical}.

Let us finally comment about the large value of the transport coefficient, $\hat q=$ 5...15 GeV$^2$/fm, obtained from the fit to inclusive particle suppression at RHIC. Several estimates indicate that this value is at least five times larger than the corresponding one for an ideal gas of quarks and gluons at the corresponding RHIC densities \cite{Eskola:2004cr}. Although no direct calculation of this quantity exists so far for a QCD medium close to the critical temperature, several interpretations are possible: i) in the interpretation of the findings at RHIC as a strongly-coupled QGP \cite{Shuryak:2005pp}, the relevant cross sections could be much larger than the perturbative ones; ii) the energy loss is sensitive to both the local energy density and present flow fields \cite{Armesto:2004pt}; iii) an additional source of energy loss is present, although, there is no clear disagreement between the findings in the formalism and different sets of experimental data. The first two possibilities open completely new ways of characterizing the medium by high-$p_t$ particle studies.

CAS is supported by the 6th Framework Programme of the European Community under the contract MEIF-CT-2005-024624.


\section*{References}
\begin{thebibliography}{99}

\bibitem{schmidt} C. Schmidt, these proceedings.

\bibitem{levkova} L. Levkova, these proceedings.

\bibitem{Matsui:1986dk}
  T.~Matsui and H.~Satz,
  Phys.\ Lett.\ B {\bf 178} (1986) 416.

\bibitem{tram} V.N. Tram these proceedings.

\bibitem{iancu} E. Iancu these proceedings.

\bibitem{lublinsky} M. Lublinsky, these proceedings.

\bibitem{gelis} F. Gelis, these proceedings.

\bibitem{Iancu:2003xm}
 For a recent review see e.g. 
  E.~Iancu and R.~Venugopalan,
 hep-ph/0303204.

\bibitem{Lappi:2006fp}
  R.~Baier, A.~H.~Mueller, D.~Schiff and D.~T.~Son,
  Phys.\ Lett.\ B {\bf 502} (2001) 51;
  T.~Lappi and L.~McLerran,
  hep-ph/0602189;
  P.~Romatschke and R.~Venugopalan,
  hep-ph/0605045.

\bibitem{Kharzeev:2000ph}
  D.~Kharzeev and M.~Nardi,
  Phys.\ Lett.\ B {\bf 507} (2001) 121.

\bibitem{loizides} C. Loizides, these proceedings.

\bibitem{Stasto:2000er}
  A.~M.~Stasto, K.~Golec-Biernat and J.~Kwiecinski,
  Phys.\ Rev.\ Lett.\  {\bf 86} (2001) 596

\bibitem{Freund:2002ux}
  A.~Freund {\it et al.}
  Phys.\ Rev.\ Lett.\  {\bf 90} (2003) 222002 

\bibitem{Armesto:2004ud}
  N.~Armesto, C.~A.~Salgado and U.~A.~Wiedemann,
  Phys.\ Rev.\ Lett.\  {\bf 94} (2005) 022002

\bibitem{Arsene:2004ux}
  I.~Arsene {\it et al.}  [BRAHMS Collaboration],
  Phys.\ Rev.\ Lett.\  {\bf 93} (2004) 242303.

\bibitem{forwardsup}
  R.~Baier, A.~Kovner and U.~A.~Wiedemann,
  Phys.\ Rev.\ D {\bf 68}, 054009 (2003);
  D.~Kharzeev, Y.~V.~Kovchegov and K.~Tuchin,
  Phys.\ Rev.\ D {\bf 68} (2003) 094013;
  J.~L.~Albacete {\it et al.}
  Phys.\ Rev.\ Lett.\  {\bf 92}, 082001 (2004).

\bibitem{Salgado:2005pr}
  For a recent review see e.g. C.~A.~Salgado,
 hep-ph/0510062 and references therein.

\bibitem{Muller:2002fa}
  B.~Muller,
  Phys.\ Rev.\ C {\bf 67} (2003) 061901;
%

\bibitem{Eskola:2004cr}
  K.~J.~Eskola {\it et al.}
  Nucl.\ Phys.\ A {\bf 747} (2005) 511.

\bibitem{Dainese:2004te}
  A.~Dainese, C.~Loizides and G.~Paic,
  Eur.\ Phys.\ J.\ C {\bf 38} (2005) 461;
  hep-ph/0511045.


\bibitem{Adcox:2001jp}
%
S.~S.~Adler {\it et al.}  [PHENIX Collaboration],
Phys.\ Rev.\ C {\bf 69} (2004) 034910;
%
%
J.~Adams {\it et al.}  [STAR Collaboration],
Phys.\ Rev.\ Lett.\  {\bf 91} (2003) 172302;
%
B.~B.~Back {\it et al.}  [PHOBOS Collaboration],
Phys.\ Lett.\ B {\bf 578} (2004) 297;
%
I.~Arsene {\it et al.}  [BRAHMS Collaboration],
Phys.\ Rev.\ Lett.\  {\bf 91} (2003) 072305.
%
  M.~Shimomura  [PHENIX],
  nucl-ex/0510023.

\bibitem{Cacciari:2005rk}
  M.~Cacciari, P.~Nason and R.~Vogt,
  Phys.\ Rev.\ Lett.\  {\bf 95} (2005) 122001

\bibitem{Armesto:2005mz}
  N.~Armesto {\it et al.}
 hep-ph/0511257.

\bibitem{Casalderrey-Solana:2006rq}
  J.~Casalderrey-Solana and D.~Teaney,
  hep-ph/0605199; and these proceedings.

\bibitem{elqm05}
S.S. Adler {\it et al.} [PHENIX] nucl-ex/0510047;
J. Bielcik [STAR] nucl-ex/0511005.

\bibitem{belikov} J. Belikov, these proceedings

\bibitem{Salgado:2003rv}
  C.~A.~Salgado and U.~A.~Wiedemann,
  Phys.\ Rev.\ Lett.\  {\bf 93} (2004) 042301

\bibitem{Adams:2006yt}
  J.~Adams  [STAR Collaboration],
 nucl-ex/0604018; 
  A.~Mischke, these proceedings.

\bibitem{Adler:2005ee}
  S.~S.~Adler {\it et al.}  [PHENIX Collaboration],
  nucl-ex/0507004.

\bibitem{Armesto:2004pt}
  N.~Armesto, C.~A.~Salgado and U.~A.~Wiedemann,
  Phys.\ Rev.\ Lett.\  {\bf 93} (2004) 242301;
%
  Phys.\ Rev.\ C {\bf 72} (2005) 064910;

\bibitem{conical}
 H.~Stoecker,
  Nucl.\ Phys.\ A {\bf 750}, 121 (2005);
J.~Casalderrey-Solana, E.~V.~Shuryak and D.~Teaney,
  hep-ph/0411315;
J. Ruppert and B.~Muller,
Phys.\ Lett.\ B {\bf 618} (2005) 123;
  V.~Koch, A.~Majumder and X.~N.~Wang,
Phys.\ Rev.\ Lett.\  {\bf 96} (2006) 172302 ;
  I.~M.~Dremin, L.~I.~Sarycheva and K.~Y.~Teplov,
  Eur.\ Phys.\ J.\ C {\bf 46} (2006) 429.

\bibitem{Shuryak:2005pp}
  See e.g. E.~Shuryak,
  hep-ph/0510123.



\end{thebibliography}
\end{document}